\documentclass[12pt,a4paper]{panl}
\usepackage{cite}
\usepackage{wrapfig}
\usepackage{graphicx}
\usepackage{amssymb}
\usepackage{amsfonts}
\usepackage{amsmath}
\usepackage{longtable}
\usepackage{rotating}
\usepackage{lscape}
\usepackage{epsfig}
\usepackage{multirow}

\originalTeX
\begin{document}

\title{Electron Identification using Machine Learning in the MPD Experiment at NICA}
\maketitle
\authors{S.\,P.\,Rode$^{a}$\footnote{E-mail: sudhirrode11@gmail.com, sudhir@jinr.ru} for the MPD collaboration}
\setcounter{footnote}{0}
\from{$^{a}$\, Veksler and Baldin Laboratory of High Energy Physics, Joint Institute for Nuclear Research,\\
Dubna, Moscow Region,
Russian Federation\\}

\begin{abstract}

We present studies of electron identification (eID) in the MPD experiment at NICA using machine learning techniques. The goal is to improve electron identification efficiency while preserving high purity, which is crucial for dielectron analyses. We compare electron identification performance between traditional cut-based approach and Machine learning. For machine learning based approach different classifiers, namely, Multi-Layer Perceptron (MLP) and Boosted Decision Tree (BDT) were trained with momentum-integrated and momentum-differential strategies using the \texttt{CERN ROOT TMVA} package.
\end{abstract}
\vspace*{6pt}


\label{sec:intro}

\section*{Introduction}
Relativistic heavy-ion collisions allow the detailed study of strongly interacting matter under extreme thermodynamic conditions. At very small baryon chemical potential, $\mu_{\rm B}\approx 0$, the transition from hadronic matter to a Quark–Gluon Plasma (QGP) is expected to be a smooth crossover, creating conditions similar to the early universe. On the other hand, at higher $\mu_{\rm B}$, one can explore the existence of a possible critical end point and first-order phase transition in the QCD phase diagram. The net baryon densities reached in these collisions are similar to those expected in neutron-star cores. The Multi-Purpose Detector (MPD)~\cite{Abgaryan2022,Abdulin2025} and Baryonic Matter at Nuclotron (BM@N) experiments at Nuclotron-based Ion Collider fAcility (NICA)~\cite{Sissakian:2009zza} are designed to study QCD matter with very high-$\mu_{\rm B}$ accessible in the designed beam energy range of $\sqrt{s_{\mathrm{NN}}} = 2.4-11$~GeV~\cite{Randrup:2006nr}.


NICA is a megascience project in Russia, approaching the final stages of construction. The commissioning of MPD, its flagship experiment, is expected in the coming months. The MPD is a barrel shaped apparatus situated within a 0.5T solenoid magnet, providing full azimuthal coverage for $|\eta| < 1.5$. Its central systems are comprised of a Time Projection Chamber (TPC), a Time-of-Flight (TOF) detector, and an Electromagnetic Calorimeter (ECal).  Two forward subsystems, the Fast Forward Detector (FFD) ($2.9 < |\eta| < 3.3$) and the Forward Hadronic Calorimeter (FHCAL) ($2 < |\eta| < 5$), extend the pseudorapidity coverage. The construction of all major subsystems is nearing completion.

The NICA collider will carry out heavy-ion collisions over a center-of-mass energy range of $\sqrt{s_{\mathrm{NN}}} = 2.4-11$~GeV, covering both fixed-target (2.4–3.5 GeV) and collider (4–11 GeV) modes. The MPD experiment is designed to operate in both modes, supporting a broad physics program across this energy range. The experimental strategy focuses on performing high-luminosity scans in collision energy and system size. Key advantage of the MPD set-up is the use of a same apparatus for all scans, the benefits of collider geometry, maximal phase-space coverage, and correlated systematic uncertainties across different systems and energies.

\section*{Track reconstruction and eID in the MPD experiment}
\begin{figure}[h!]
\centering

\includegraphics[scale=0.25]{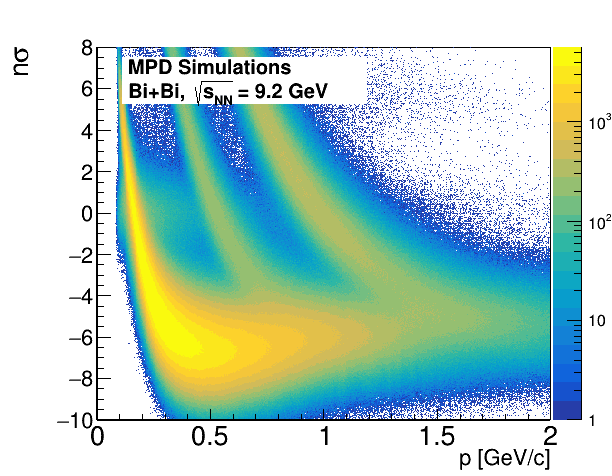}
\includegraphics[scale=0.25]{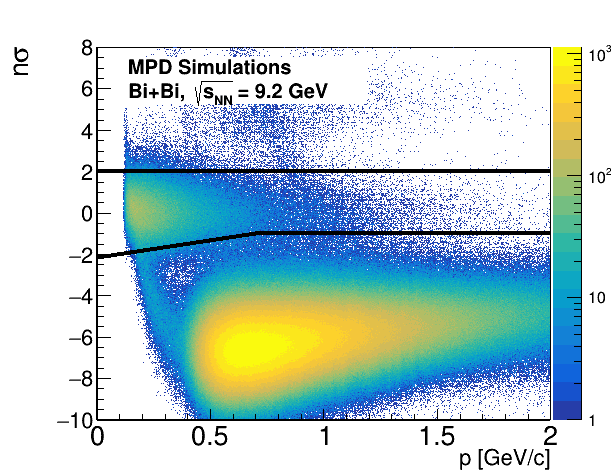}\\
\includegraphics[scale=0.25]{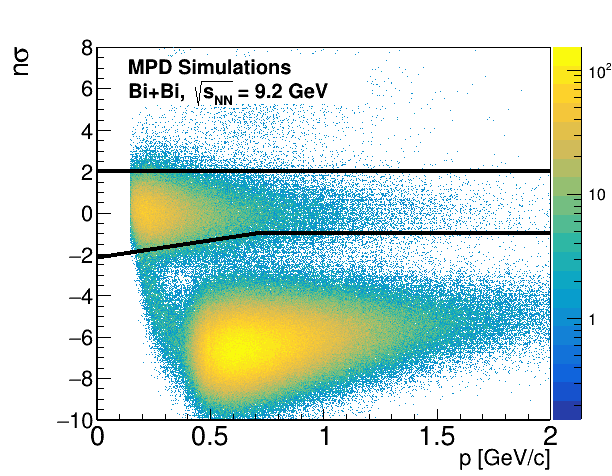}

\caption{The n$\sigma$ deviations of dE/dX of charged tracks in the TPC (top left) with additon of TOF PID (tof right) and TOF + ECal PID (bottom) as a function of linear momentum.}
\label{f1}
\end{figure}
Particle identification (PID) is a fundamental step in physics analyses in heavy-ion collisions. The MPD detector provides strong track reconstruction and PID capabilities. The TPC detector measures the average ionization energy loss, $\langle {\rm d}E/{\rm d}x\rangle$, and momentum with typical resolutions of about 6--7\% (dE/dx) and 1--3\% (momentum). The TPC alone is, however, insufficient to reach a high purity electron sample. The TOF detector complements TPC with the help of time-of-flight information and improves the purity, and the ECal further enhances electron-hadron separation using measured deposited energy and full momentum, $E/p$ ratio, as well as time-of-flight measurement. Fig.~\ref{f1} depicts the TPC detector response in terms of number of sigmas without TOF and ECal (left panel), with TOF PID (right panel) and additionally, ECal PID (bottom panel).

\begin{figure}[htbp]
  \centering
  \includegraphics[scale=0.18]{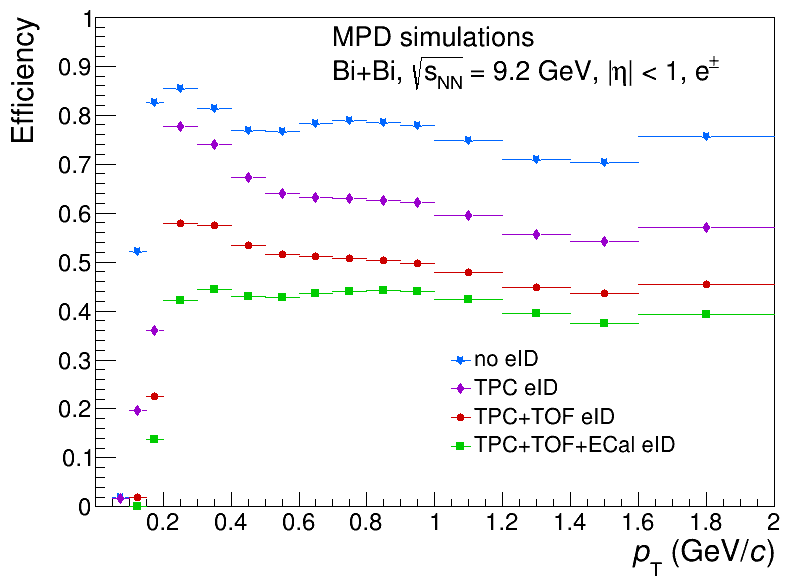}
  \includegraphics[scale=0.18]{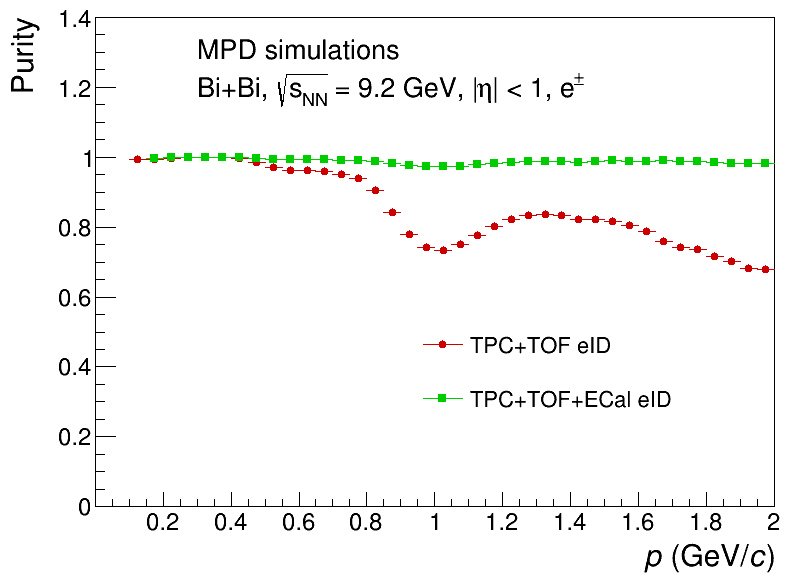}
  \caption{Electron reconstruction efficiency and purity using traditional 1D cuts.}
\label{f2}
\end{figure}

Electron identification (eID) using traditional one-dimensional selection cuts is characterized by efficiency and purity as shown in Fig.~\ref{f2}. With TOF the purity reaches approximately 80\% while efficiency is around 50\%, and adding ECal can reach near 100\% purity at the cost of a reduced efficiency of $\approx40\%$. As shown in the left plot of Fig.~\ref{f2}, the efficiency drops with each selection cut. This motivates us to find an alternative eID approach to reduce this efficiency loss without compromising on the purity.

\section*{eID using Machine Learning}
Machine learning has been used to perform particle identification tasks in many physics analyses in heavy-ion collisions~\cite{Papoyan:2025kdk,Tolkachev:2023nwg,Papoyan:2023sap,Graczykowski:2022zae,Karwowska:2024xqy}
, and it offers a promising approach to improve electron identification performance in MPD. The ML-based approach used here aims to increase electron identification efficiency while keeping purity high, which can provide assistance in improving the signal-to-background ratio and statistical significance in dielectron analyses. For this study, the TMVA package from CERN ROOT framework is employed~\cite{Brun:1997pa,Therhaag:2010zz,TMVA:2007ngy}.
\begin{figure}[h!]
\centering
    \includegraphics[scale=0.15]{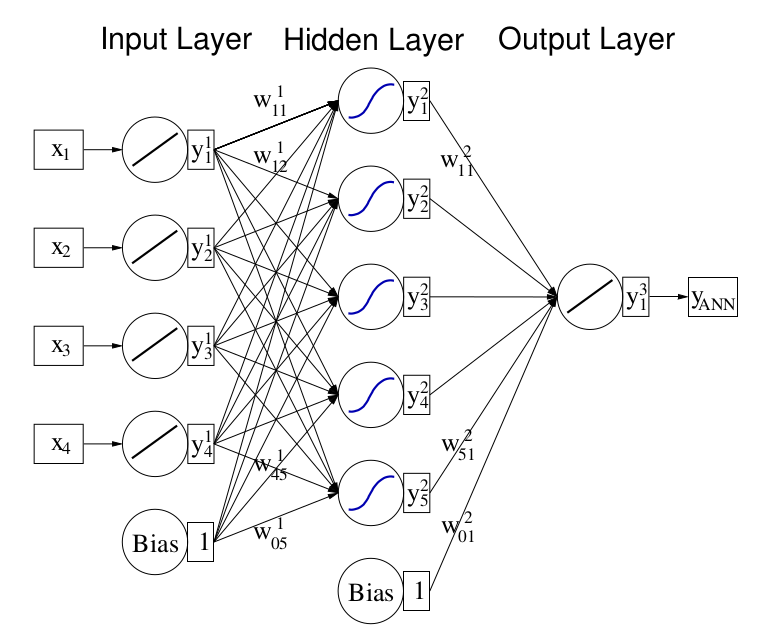}
    \includegraphics[scale=0.4]{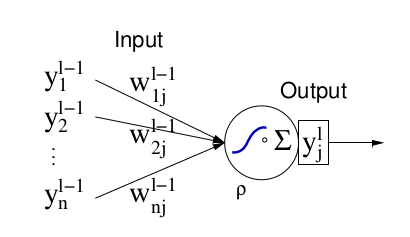}
    \caption{MLP (neural network) with one hidden layer (left). Neuron response function mapping input onto output (right).}
\label{f3}    
    \includegraphics[scale=0.15]{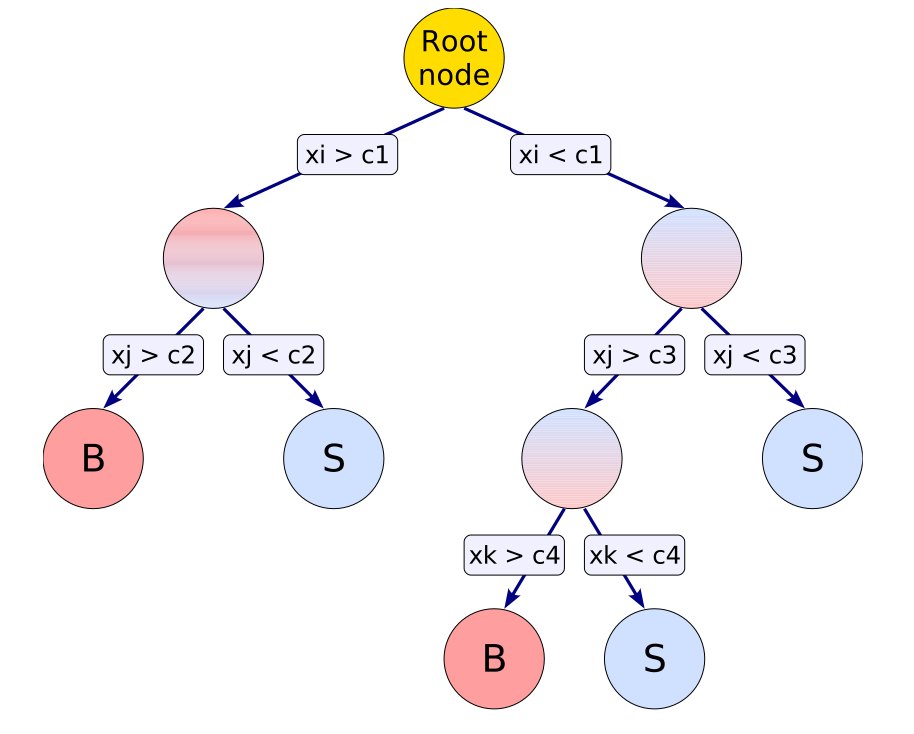}
    \caption{Schematic view of a decision tree. Starts from the root node with a sequence of binary splits using the discriminating variables.}
\label{f4}
\end{figure}
The general procedure followed in this work is:
\begin{enumerate}
  \item Preparation of the data samples involves full detector simulation of Bi+Bi collisions at $\sqrt{s_{NN}}$ = 9.2 GeV obtained using the Ultra-relativistic Quantum Molecular Dynamics (UrQMD) event generator~\cite{Bass:1998ca,Bleicher:1999xi} within the \texttt{MPDROOT} framework. The dataset of reconstructed charged tracks is splitted into independent training and validation (testing) sets. Typical splits used here are 50\% for training and 50\% for validation while preserving the actual signal/background proportions.
  \item Each charged track has different discriminating variables that provide separation between electrons (signal) and hadrons (background). The input variables include full momentum ($p$), number of hits in the TPC, energy loss in the TPC ($dE/dx$), velocity of the particle measured in the TOF and ECal ($\beta$), energy-to-momentum ratio in the ECal ($E/p$), distance of closest approach ($\rm DCA_{x,z}$), $\chi^{2}$ of the track at the vertex, azimuthal angle ($\phi$), and pseudorapidity ($\eta$). These variables were chosen to exploit detector responses that are different for electrons and hadrons.

  \item The classifiers from the TMVA package, in this case, Multi-Layer Perceptron (MLP) neural network and Boosted Decision Trees (BDT) are trained on the training sample and for performance validation the independent testing sample is used.
\end{enumerate}

MLP and BDT are based on different principles. MLP is a feed-forward neural network in which neurons, responsible for learning patterns in data, are arranged in successive layers. Each layer is fully connected to the next, allowing information to propagate through weighted connections. The neuron response functions, also known as activation functions, provide non-linear mappings from inputs to outputs, enabling the network to capture complex correlations in the data. During training, the weights are iteratively optimized to minimize classification errors. On the other hand, BDT is an ensemble method that constructs a sequence of simple decision trees to improve classification performance. Each individual tree applies a set of step-by-step ``if--then'' rules to partition the feature space into regions with enhanced class purity. By focusing sequentially on misclassified events, the boosting algorithm transforms weak learners into a strong predictor through error correction. The final decision is obtained from a weighted combination of the individual trees, leading to improved accuracy and robustness. The schematics of both classifiers are shown in Figs.~\ref{f3} and~\ref{f4}. For more details about the classifiers, please see Ref.~\cite{TMVA:2007ngy}.

\section*{Results and Discussion}

\begin{figure}[h!]
\centering
\includegraphics[scale=0.25]{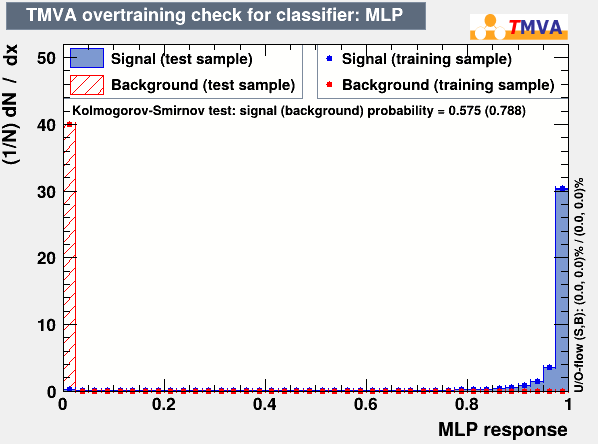}
\includegraphics[scale=0.25]{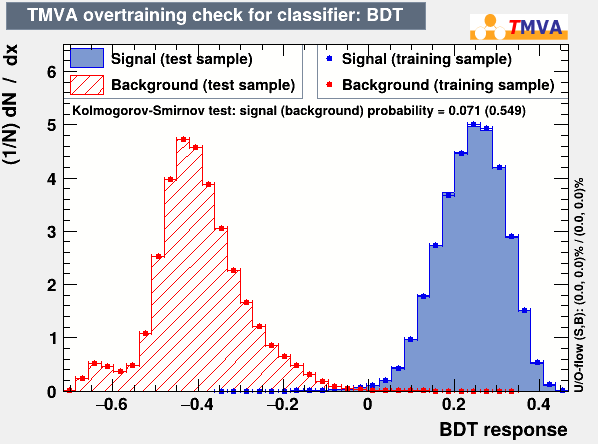}
  \caption{Classifer response and Overtraining tests for momentum-integrated training: MLP (left) and BDT (right).}
\label{f6}
\end{figure}

\begin{figure}[h!]
\centering
\includegraphics[scale=0.17]{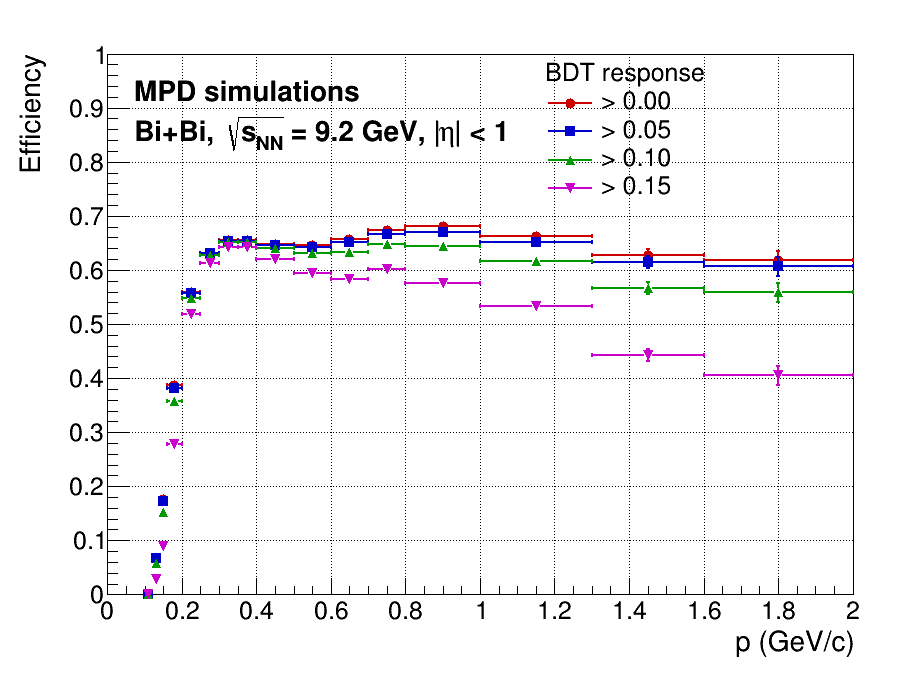}
\includegraphics[scale=0.17]{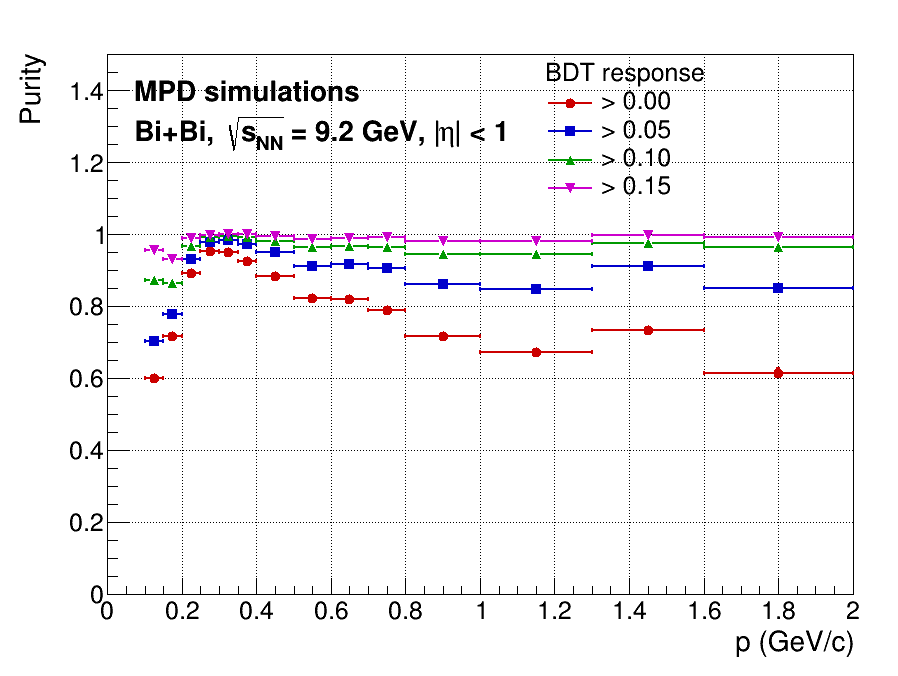}\\
\includegraphics[scale=0.17]{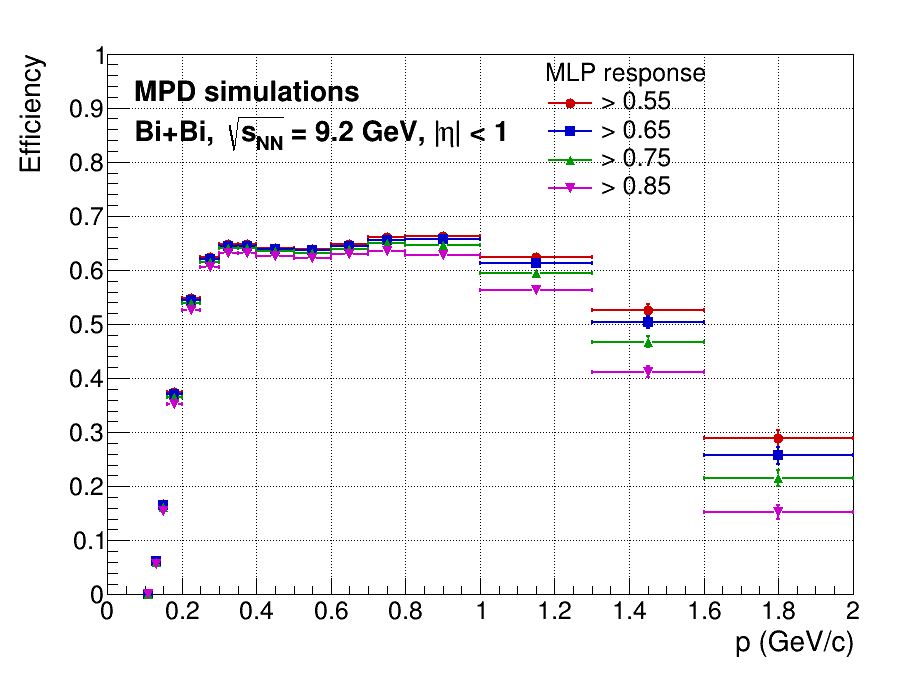}
\includegraphics[scale=0.17]{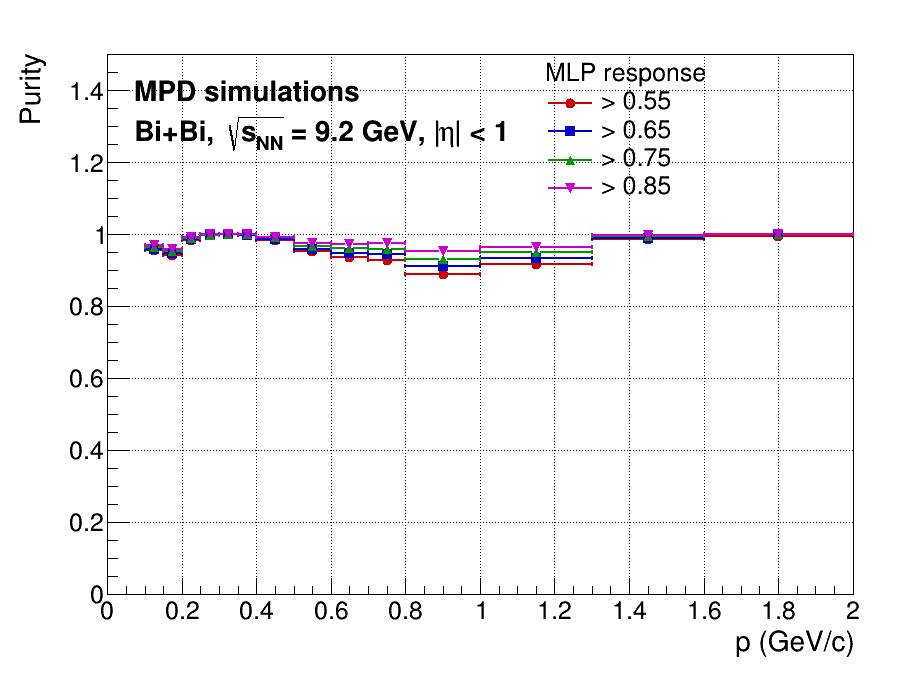}
\caption{Electron reconstruction efficiency (left) and purity (right) for BDT (top) and MLP (bottom) with Momentum-integrated training.}
\label{f7}
\end{figure}

We first performed momentum-integrated training where a classifier is trained on the whole sample with full momentum interval. To avoid overtraining or overfitting, the so-called Kolmogorov-Smirnov test was performed. As a rule of thumb, it is recommended to try to reduce overtraining if $p$-value obtained in this test is, $p$ $<$ 0.01, especially if the separation is visibly poorer for the testing samples than for the training samples. However, no significant overtraining was observed for the configurations used as denoted by $p$ value in Fig.~\ref{f6}. As a part of performace validation, electron identification efficiency and purity as functions of momentum were estimated and it was found that both MLP and BDT lose discrimination power at high momentum as shwon in Fig.~\ref{f7}

To overcome this efficiency loss at high momenta, we performed momentum-differential training where classifiers are trained separately in different momentum intervals, namely (0.0-0.3 GeV/c), (0.3-0.6 GeV/c), (0.6-0.9 GeV/c), (0.9-1.2 GeV/c), (1.2-1.6 GeV/c) and (1.6-5.0 GeV/c). Training the sample in small momentum intervals can help the classifiers to achieve better separation at high momentum and allows the classifiers to adapt to changing discrimination power of the input variables.

\begin{figure}[h!]
\centering

\includegraphics[scale=0.18]{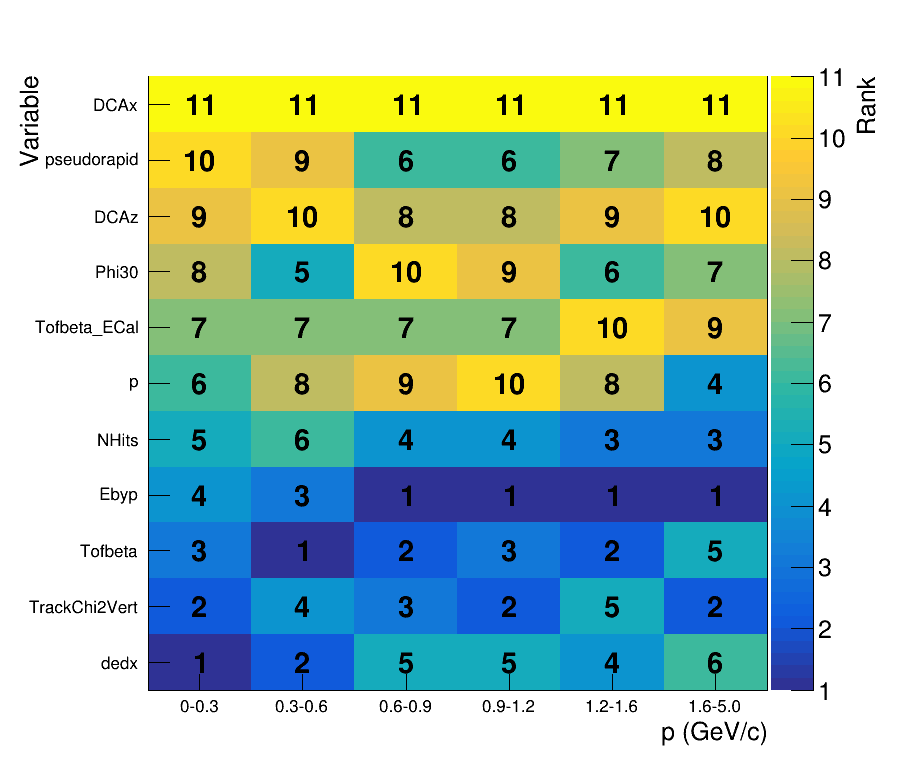}
\includegraphics[scale=0.18]{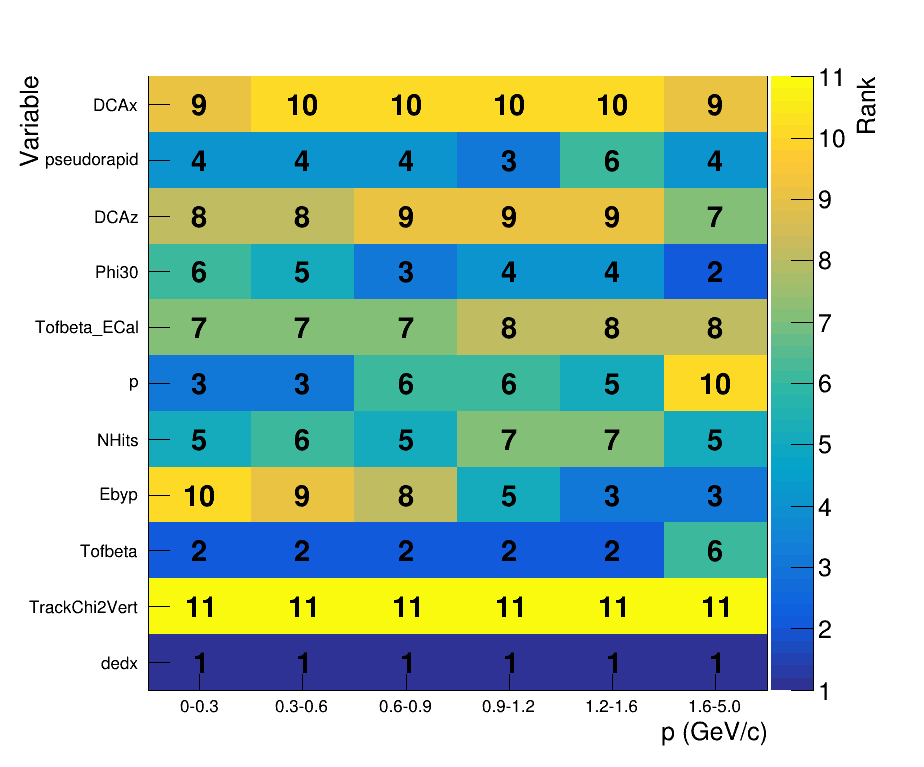}
\caption{The varibles vs. linear momentum intervals depicting input variables importance rankings for MLP (left) and BDT (right).}
\label{f8}
\end{figure}

\begin{figure}[h!]
\centering
\includegraphics[scale=0.17]{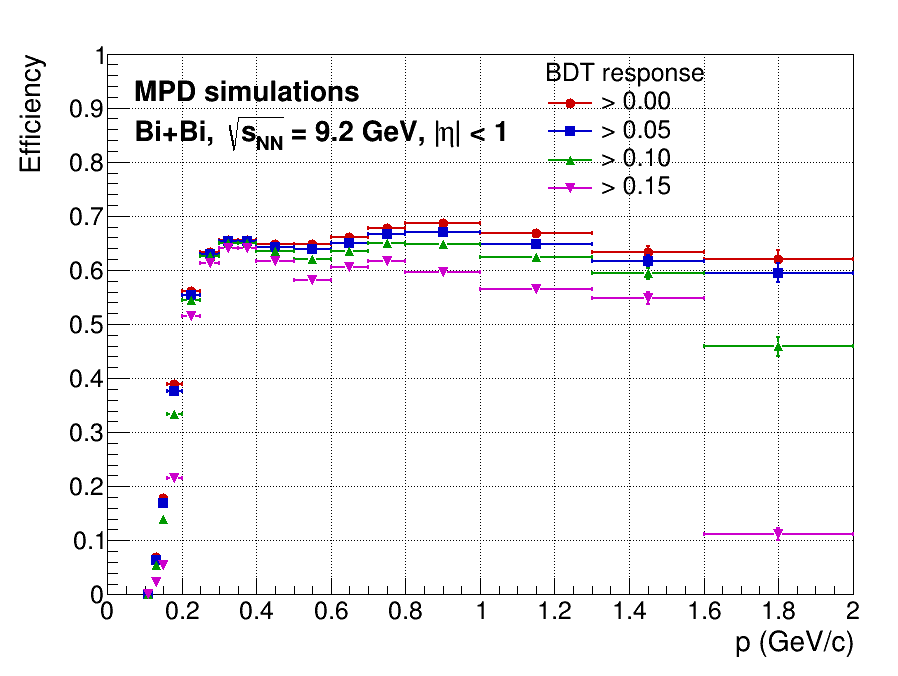}
\includegraphics[scale=0.17]{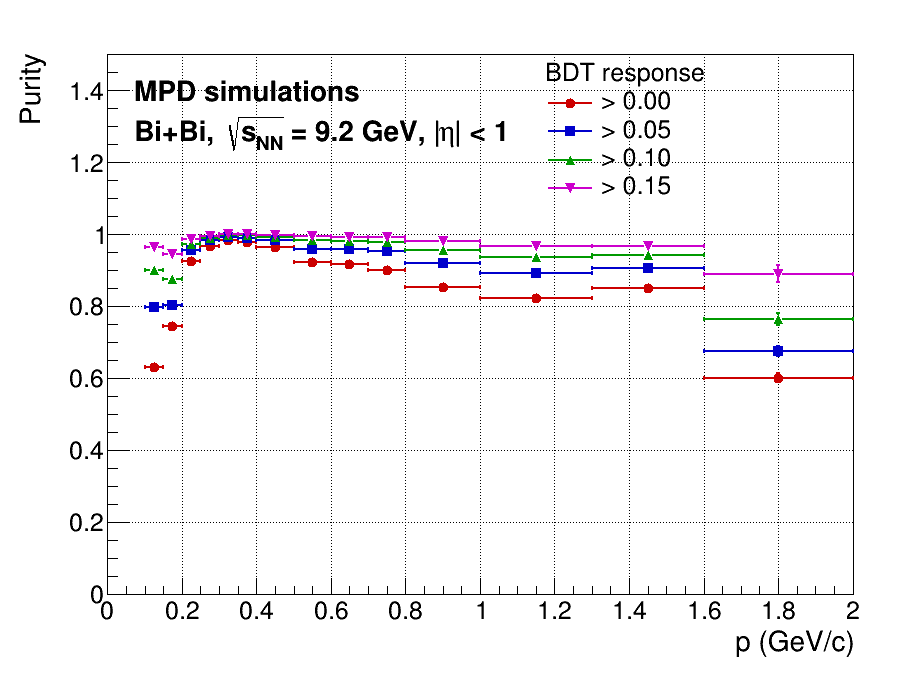}\\
\includegraphics[scale=0.17]{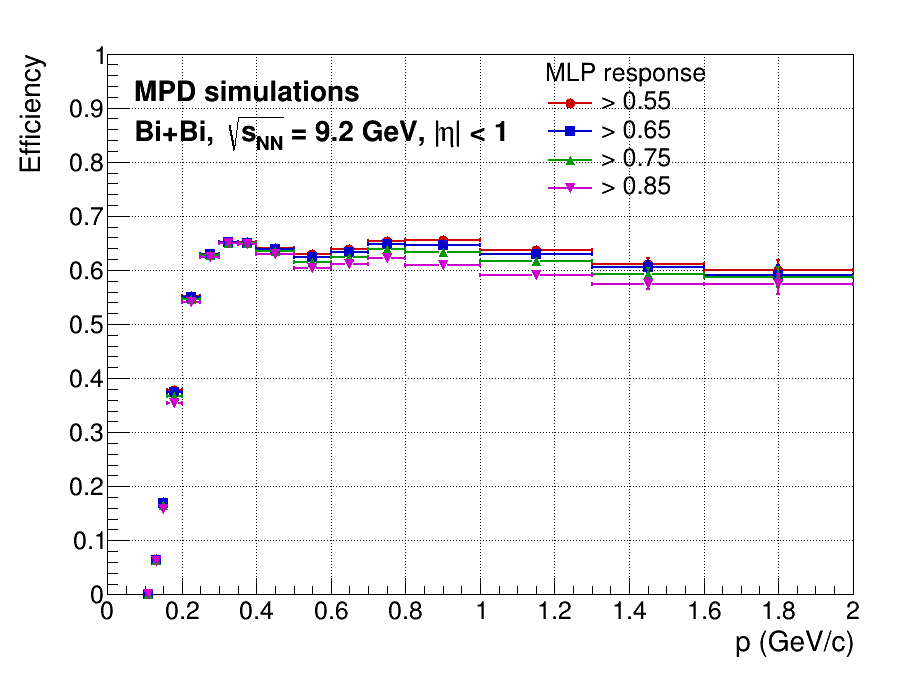}
\includegraphics[scale=0.17]{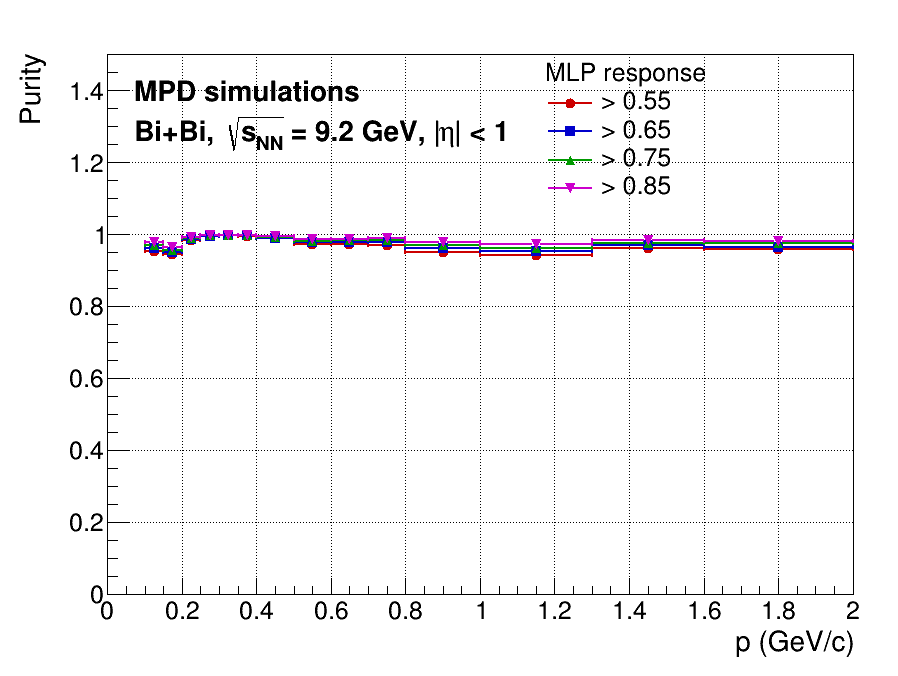}
\caption{Electron reconstruction efficiency (left) and purity (right) for BDT (top) and MLP (bottom) with Momentum-differential training.}
\label{f9}
\end{figure}

\begin{figure}[h!]
\centering
\includegraphics[scale=0.17]{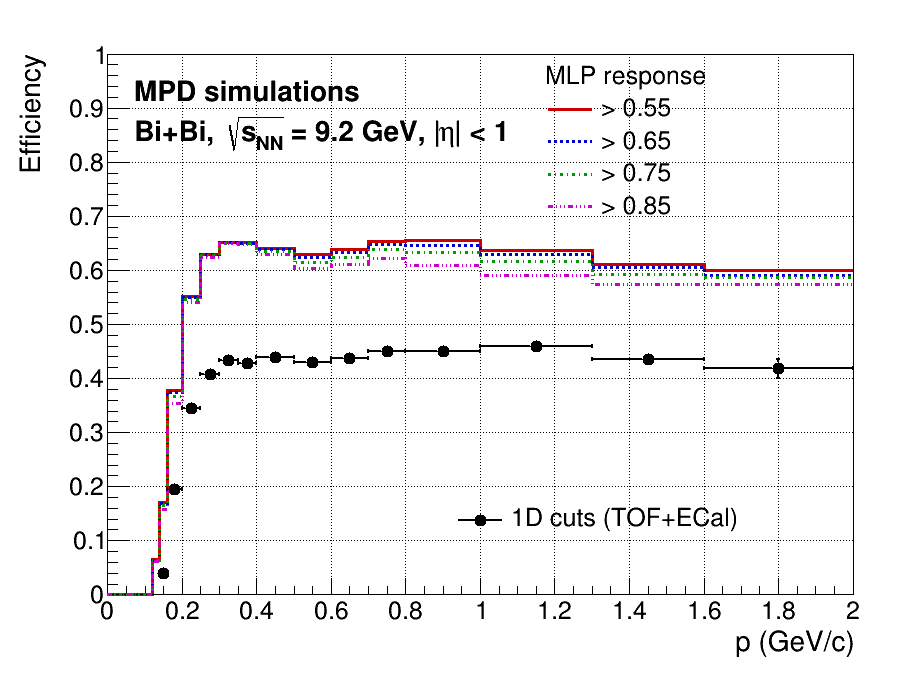}
\includegraphics[scale=0.17]{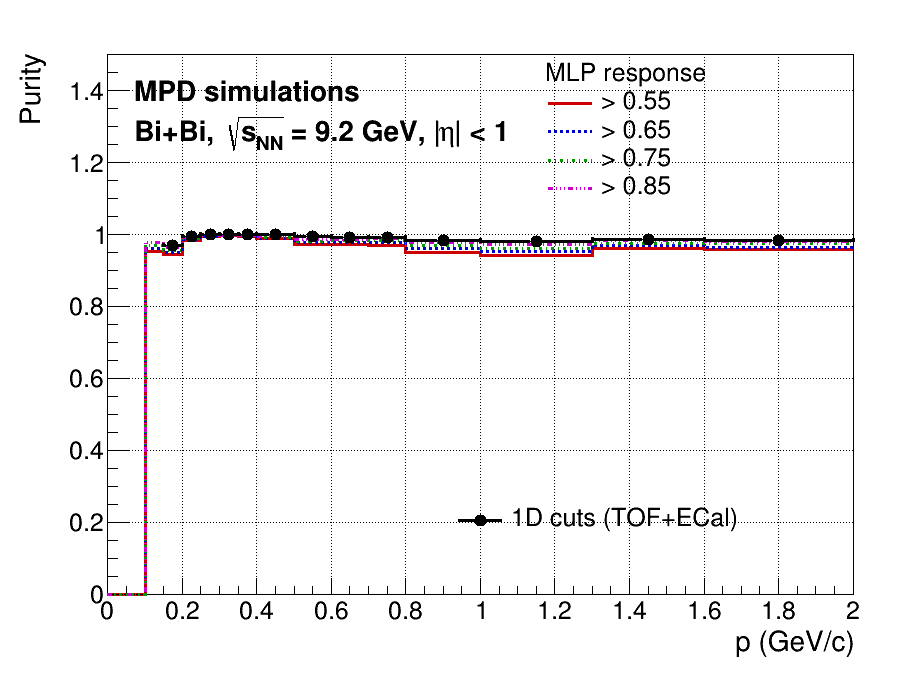}
\caption{Comparison between electron efficiency and purity obtained using 1D cuts and MLP classifier with momentum-differential training.}
\label{f10}
\end{figure}

In Fig.~\ref{f8}, variable importance rankings in different momentum ranges are shown for MLP (left plot) and BDT (right). In the MLP neural network, variable ranking is based on the sum of the squared weights connecting an input neuron to the neurons in the first hidden layer. As for BDT, variable ranking is based on how often a variable is used to split nodes, with each split weighted by its separation gain and the number of events it contains. The plots suggest classifier-dependent sensitivities to the input variables in different momentum ranges. MLP and BDT show different ranking of variable importance, and at higher momenta the classifiers tend to rely on different combinations of the input observables. For instance, MLP seem to utilize E/p more at high momenta where ECal is effective for electron-hadron separation, whereas, BDT utilize dE/dx at every interval. Consequently, the MLP maintains more stable efficiency and purity across momentum than the BDT in these studies as shown in Fig.~\ref{f9}.

A direct comparison between the traditional 1D-cut method and the MLP-based momentum differential training approach, as shown in Fig.~\ref{f10}, demonstrates a clear improvement in electron identification efficiency using the MLP while maintaining high purity. The efficiency is improved by about 50\% while purity remains nearly 100\%. Machine learning methods, therefore, provide a clear advantage for eID in MPD analyses. As a result, it is foreseen that the achieved improvement in the single electron identification could provide about factor 2 enhancement in the dielectron signal measurement in the dielectron analysis. 

The study presented here was conducted using a detailed simulation of nominal detector design. The primary aim of the present work was to establish whether a significant performance gain is achievable with an ML-based approach. A dedicated study to evaluate the stability of the model under detector variations has not been performed as it is beyond the present scope.

\section*{Conclusions}
All MPD sub-systems are in final stages of construction and are expected to be ready for commissioning in the coming months. MPD operations are expected to begin in 2026. In general, particle identification is a key step in physics analyses as well as in the context of the MPD physics program. Traditional cut-based particle identification methods suffer significant efficiency loss, motivating the use of machine learning techniques. Both MLP and BDT classifiers improve electron identification performance compared to 1D cuts; however, the MLP performs better in the high-momentum region, where momentum-differential training helps to achieve uniform efficiency and purity. These improvements are important for dielectron and electromagnetic-probe analyses at NICA.

\bigskip
\noindent\textbf{Acknowledgements.} The author thanks the MPD collaboration for their encouragement and suggestions.

\bigskip
\noindent\textbf{Funding:}
This work was supported by ongoing institutional funding. No additional grants to carry out or direct this particular research were obtained.

\bigskip
\noindent\textbf{Conflict Of Interest:}
The author of this work declares that he has no conflicts of interest.


\begin{thebibliography}{1}
\def\selectlanguageifdefined#1{
\expandafter\ifx\csname date#1\endcsname\relax
\else\selectlanguage{#1}\fi}
\providecommand*{\href}[2]{{\small #2}}
\providecommand*{\url}[1]{{\small #1}}
\providecommand*{\BibUrl}[1]{\url{#1}}
\providecommand{\BibAnnote}[1]{}
\providecommand*{\BibEmph}[1]{\emph{#1}}
\ProvideTextCommandDefault{\cyrdash}{\hbox to.8em{--\hss--}}
\providecommand*{\BibDash}{\ifdim\lastskip>0pt\unskip\nobreak\hskip.2em\fi
\cyrdash\hskip.2em\ignorespaces}

\bibitem{Abgaryan2022}
\selectlanguageifdefined{english}
\BibEmph{Abgaryan V. et al.} [MPD] {Eur. Phys. J. A}~//
\href{http://dx.doi.org/10.1140/epja/s10050-022-00792-w}{Eur. Phys. J. A}
\BibDash
\newblock 2022. \BibDash
\newblock V.~58. \BibDash
\newblock P.~140. \BibDash

\bibitem{Abdulin2025}
\selectlanguageifdefined{english}
\BibEmph{Abdulin R. et al.} [MPD] {Rev. Mex. Fis.}~//
\href{http://dx.doi.org/10.31349/RevMexFis.71.041201}{Rev. Mex. Fis.}
\BibDash
\newblock 2025. \BibDash
\newblock V.~71, no.~4. \BibDash
\newblock P.~041201. \BibDash
\bibitem{Sissakian:2009zza}
\selectlanguageifdefined{english}
\BibEmph{Sissakian A.~N. et al.} [NICA] {J. Phys. G}~//
\href{http://dx.doi.org/10.1088/0954-3899/36/6/064069}{Phys. Part. Nucl. Lett.}
\BibDash
\newblock 2009. \BibDash
\newblock V.~36. \BibDash
\newblock P.~064069. \BibDash

\bibitem{Randrup:2006nr}
\selectlanguageifdefined{english}
\BibEmph{Randrup J. and Cleymans J.} {Phys. Rev. C}~//
\href{http://dx.doi.org/10.1103/PhysRevC.74.047901}{Phys. Rev. C}
\BibDash
\newblock 2006. \BibDash
\newblock V.~74. \BibDash
\newblock P.~047901. \BibDash

\bibitem{Papoyan:2025kdk}
\selectlanguageifdefined{english}
\BibEmph{Papoyan V., Aparin A., Ayriyan A., Grigorian H. and Korobitsin A.} {Phys. Part. Nucl. Lett.}~//
\href{http://dx.doi.org/10.1134/S1547477125700256}{Phys. Part. Nucl. Lett.}
\BibDash
\newblock 2025. \BibDash
\newblock V.~22, no.~3. \BibDash
\newblock P.~622--628. \BibDash

\bibitem{Tolkachev:2023nwg}
\selectlanguageifdefined{english}
\BibEmph{Tolkachev G., Korobitsin A. and Aparin A.} {Phys. Atom. Nucl.}~//
\href{http://dx.doi.org/10.1134/S1063778823050381}{Phys. Atom. Nucl.}
\BibDash
\newblock 2023. \BibDash
\newblock V.~86, no.~5. \BibDash
\newblock P.~845--849. \BibDash

\bibitem{Papoyan:2023sap}
\selectlanguageifdefined{english}
\BibEmph{Papoyan V., Aparin A., Ayriyan A., Grigorian H., Korobitsin A. and Mudrokh A.} {Phys. Atom. Nucl.}~//
\href{http://dx.doi.org/10.1134/S1063778823050332}{Phys. Atom. Nucl.}
\BibDash
\newblock 2023. \BibDash
\newblock V.~86, no.~5. \BibDash
\newblock P.~869--873. \BibDash


\bibitem{Graczykowski:2022zae}
\selectlanguageifdefined{english}
\BibEmph{Graczykowski {\L}.~K. et al.} [ALICE] {JINST}~//
\href{http://dx.doi.org/10.1088/1748-0221/17/07/C07016}{JINST}
\BibDash
\newblock 2022. \BibDash
\newblock V.~17, no.~07. \BibDash
\newblock P.~C07016. \BibDash
\newblock arXiv:2204.06900~[nucl-ex].

\bibitem{Karwowska:2024xqy}
\selectlanguageifdefined{english}
\BibEmph{Karwowska M. et al.} [ALICE] {JINST}~//
\href{http://dx.doi.org/10.1088/1748-0221/19/07/C07013}{JINST}
\BibDash
\newblock 2024. \BibDash
\newblock V.~19, no.~07. \BibDash
\newblock P.~C07013. \BibDash
\newblock arXiv:2403.17436~[hep-ex].


\bibitem{Brun:1997pa}
\selectlanguageifdefined{english}
\BibEmph{Brun R. and Rademakers F.} {Nucl. Instrum. Meth. A}~//
\href{http://dx.doi.org/10.1016/s0168-9002(97)00048-x}{Nucl. Instrum. Meth. A}
\BibDash
\newblock 1997. \BibDash
\newblock V.~389, no.~1-2. \BibDash
\newblock P.~81--86. \BibDash

\bibitem{Therhaag:2010zz}
\selectlanguageifdefined{english}
\BibEmph{Therhaag J.} {PoS}~//
\href{http://dx.doi.org/10.22323/1.120.0510}{PoS}
\BibDash
\newblock 2010. \BibDash
\newblock V.~ICHEP2010. \BibDash
\newblock P.~510. \BibDash



\bibitem{TMVA:2007ngy}
\selectlanguageifdefined{english}
\BibEmph{Hocker A. et al.} [TMVA] {arXiv:physics/0703039 [physics.data-an]}~//
\href{http://arxiv.org/abs/physics/0703039}{arXiv}
\BibDash
\newblock 2007. \BibDash

\bibitem{Bass:1998ca}
\selectlanguageifdefined{english}
\BibEmph{Bass S.~A. et al.} {Prog. Part. Nucl. Phys.}~//
\href{http://dx.doi.org/10.1016/S0146-6410(98)00058-1}{Prog. Part. Nucl. Phys.}
\BibDash
\newblock 1998. \BibDash
\newblock V.~41. \BibDash
\newblock P.~255--369. \BibDash

\bibitem{Bleicher:1999xi}
\selectlanguageifdefined{english}
\BibEmph{Bleicher M. et al.} {J. Phys. G}~//
\href{http://dx.doi.org/10.1088/0954-3899/25/9/308}{J. Phys. G}
\BibDash
\newblock 1999. \BibDash
\newblock V.~25. \BibDash
\newblock P.~1859--1896. \BibDash

\end{thebibliography}

\end{document}